\documentclass{aa}  

\usepackage{graphicx}
\usepackage{txfonts}
\usepackage{colortbl}
\usepackage{hyperref}
\hypersetup{colorlinks=true,linkcolor=[rgb]{1.,0.2,0.2},citecolor=[rgb]{0.1,0.1,1.},filecolor=[rgb]{0.7,0.2,0.2},urlcolor=[rgb]{0.7,0.2,0.2}}
\usepackage{pifont}
\usepackage{adjustbox}
\newcommand{\msun}{\,\mathrm{M_\odot}}

\def \Fiducial{\textit{Fiducial}}
\def \Wandering{\textit{Off-center}}

\def \LGalaxies{\texttt{L-Galaxies}\textit{BH}\,}
\def \LGalaxiesHenriques{\texttt{L-Galaxies}\,}

\usepackage{ulem}

\def \msun{\,\rm M_\odot}

\usepackage{color}
\definecolor{myorange}{rgb}{0.8, 0.3, 0.0}

\definecolor{mygreen}{rgb}{0.0, 0.398, 0.0}

\begin{document} 

 \title{Off-center black hole seed formation?\\
   Implications for high and low redshift massive black holes}
   \titlerunning{Off-center black hole seed formation?}
   \author{David Izquierdo-Villalba$^{*1,2}$ \and Daniele Spinoso$^{3}$ \and Marta Volonteri$^4$ \and Monica Colpi$^{1,2}$ \and \\ Alberto Sesana$^{1,2,5}$ \and Silvia Bonoli$^{6,7}$}
   \institute{
     $^{1}$ Dipartimento di Fisica ``G. Occhialini'', Universit\`{a} degli Studi di Milano-Bicocca, Piazza della Scienza 3, I-20126 Milano, Italy\\
     $^{2}$ INFN, Sezione di Milano-Bicocca, Piazza della Scienza 3, 20126 Milano, Italy\\ 
     $^{3}$ Como Lake Center for AstroPhysics,  University of Insubria, 22100, Como, Italy\\ 
     $^{4}$ Institut d’Astrophysique de Paris, Sorbonne Université, CNRS, UMR 7095, 98 bis bd Arago, 75014 Paris, France\\
     $^{5}$ INAF - Osservatorio Astronomico di Cagliari, via della Scienza 5, 09047 Selargius (CA), Italy\\
     $^{6}$ Donostia International Physics Centre (DIPC), Paseo Manuel de Lardizabal 4, 20018 Donostia-San Sebastian, Spain\\
     $^{7}$ IKERBASQUE, Basque Foundation for Science, E-48013, Bilbao, Spain \\ \\
   \email{david.izquierdovillalba@unimib.it}}

   \date{Received; accepted}

  \abstract {
    Recent studies show that light seeds of black holes, which grow into massive black holes (MBHs) over time, often struggle to remain at the centers of their birthplaces in high-redshift galaxies, limiting their ability to accrete gas and merge with other black holes. In this work, we investigate how off-center formation of the first seeds affects the evolution of the MBH and massive black hole binary (MBHB) populations over cosmic history. To this end, we use the \texttt{L-Galaxies}{\it BH} semi-analytical model, which includes multiple seed formation mechanisms, with light Population III remnants being the most significant contributors. To incorporate off-center formation, we modify the model to track the initial seed location, the sinking timescales toward the galactic center, and any growth during this phase. The results indicate that seed formation occurring away from the galactic center has a negligible impact on the MBH population at $z\,{<}\,1$, but causes significant differences at higher redshifts. Particularly, the abundance of ${>}\,10^5\,\msun$ MBHs at $z\,{>}\,4$ can be up to 2-10 times smaller compared to a nuclear seed formation model. Quasar luminosity functions with $\rm L_{bol} \,{>}\, 10^{44}\,\rm erg/s$ are similarly affected, although they still align with observational constraints. The off-centre formation also alters the galaxy-MBH mass relation. At $z \,{>}\, 5$, the amplitude of the relation can be up to 2 dex smaller than in nuclear seed models. These differences fade by $z \,{\sim}\, 2$ for galaxies ${>}\,10^{11}\,\msun$, and by $z\,{=}\,0$ for smaller galaxies. Notably, the overmassive MBH population recently unveiled by JWST is still present in the model, suggesting they can form independently of the seed dynamics. Finally, the merging rate of MBHs within LISA sensitivity band is strongly impacted. Specifically, there is a suppression of events at high-$z$ and an enhancement at low-$z$. 

   }
   \keywords{Methods: numerical --- quasars: supermassive black holes -- Gravitational waves -- }
    \authorrunning{Izquierdo-Villalba et al}
   \maketitle
%

%

\section{Introduction}

Our understanding of massive black holes (MBHs) has greatly  advanced since the discovery of the first quasar \citep{Schmidt1963}. Observations in the nearby Universe have confirmed the presence of MBHs at the center of most massive galaxies. Furthermore, these studies have underscored the importance of gas accretion in powering quasars and revealed the connection between MBH masses and the properties of their host galaxies 
\citep{Soltan1982, Haehnelt1993, HaringRix2004, Marconi2004, Kormendy2013, Savorgnan2016}. 
These investigations have been extended to higher redshifts thanks to the discovery of hyper-luminous quasars (${>}\,10^{47}\, \rm erg/s$ with ${>}\,10^8\, \msun$ at $z\,{>}\,7$, \citealt{Wu2015, Baniados2018, Yang2020}) and the large population of candidate AGNs ($\rm 10^{44\,{-}\,46}\, erg/s$ with $\,{>}\,10^6\, \msun$ at $z\,{>}\,4$) recent unveiled by the James Webb Space Telescope \citep[JWST,][]{Maiolino2023, Matthee2024, Harikane2023, Kokorev2024, Geris2025}.

Determining the abundance and properties of MBHs in the early Universe is essential for uncovering their origins. A census of the masses and luminosities of high-$z$ quasars becomes crucial to offer key constraints on the processes behind MBH formation, commonly referred to as \textit{MBH seeding} (see \citealt{Volonteri2010,inayoshi_visbal_haiman2020}). The formation of the first MBHs remains uncertain, with several competing hypotheses under investigation. One leading scenario involves the formation of light MBH seeds (${\sim}\,100\, \msun$). These seeds are thought to be remnants of Population III (PopIII) stars formed in halos of $10^6\, \msun$ at $z\,{>}\,15$ \citep{MadauRees2001,Schneider2002,Schaerer2002,BrommLarson2004,Schneider2006,Latif2015,Haemmerle2018}. Recent studies suggest that these light seeds can evolve into the bright, rare AGNs observed at $z\,{>}\,6$ as long as they experience episodes of super-Eddington accretion \citep{Pezzulli2017} or if they form as early as $z\,{>}\,35$ and grow with at a constant Eddington rate during ${\sim}\,60\%$ of their lifetime \citep{MadauRees2001,Tanaka2009}. Other scenarios suggest that intermediate-mass (${\sim}\,10^3 \, \msun$) and heavy (${\sim}\,10^{5-6} \, \msun$) MBH seeds may have formed at $z\,{>}\,7$ through stellar (black hole) runaway collisions in dense star clusters or  direct collapse of gas clouds under rather particular conditions to avoid gas-cooling and fragmentation \citep{Ebisuzaki2001,Begelman2006,LodatoNatarajan2006,DevecchiVolonteri2009,Devecchi2012,Regan2014,Lupi2014,Latif2015b,Valiante2017,Reinoso2018,ChonOmukai2020,Sassano2021,Rantala2024}. Although these alternative scenarios allow relaxing  the need for super-Eddington or nearly continuous Eddington-limited accretion \citep[see also][]{Lupi2024}, the assembly of a population of ${>}\,10^6\, \msun$ at $6\,{<}\,z\,{<}\,20$  requires specific conditions to be fulfilled, and certain level of sustained mass accretion remains necessary.

Despite the various MBH seeding hypotheses, gas accretion onto MBH seeds remains a fundamental process for explaining the existence of ${>}\,10^6\, \msun$ MBHs at $z\,{>}\,6$. How seeds manage to sustain their growth is still uncertain and controversial. 
Using hydrodynamical simulations, \cite{Mehta2024} showed that, in the absence of supernova feedback, light seeds can remain at the center of their natal gas clumps and grow efficiently over just a few Myr. \cite{Shi2024} reported complementary results, indicating that while feedback processes hinder their growth, light seeds can still experience short phases of very high accretion rates. Conversely, other studies focusing on MBH dynamics have questioned whether light seeds can sustain significant growth phases. For instance, using high-resolution simulations, \cite{Smith2018} tracked the growth of thousands of light seeds (${\lesssim}\,100\,\msun$) formed within mini-halos in the early Universe. Their findings showed that these MBHs are scattered throughout their host galaxies, preventing them from residing in the nuclear clumps of cool, dense gas. As a result, light seeds rarely undergo significant periods of growth, limiting the formation of an intermediate population of MBHs in the early Universe. Simulations by \cite{Pfister2019} on the dynamics of high-$z$ seeds showed that light MBH seeds (${<}\,10^5\,\msun$) are prone to starvation, as they are more easily scattered within their host galaxy due to irregularities of the gas/stellar potential, remaining far from dense gas regions limiting their growth (see \citealt{Bellovary2019} for comparable findings in low-$z$ dwarf galaxies). The difficulties experienced by light MBH seeds in their migration towards galactic centers were also confirmed by  the simulations of \cite{Ma2021}, who showed that the chaotic, rapidly evolving, clumpy and bursty nature of high-redshift galaxies hinders the possibility of low-mass seeds to sink efficiently from the outskirts to the central regions. Several studies have also shown that MBH seeds can overcome these difficulties and stabilise at the galactic center (experiencing significant growth) as long as they form within and co-evolve with nuclear star clusters (NSCs). Specifically, the mass of these NSCs is typically sufficient to resist external perturbations and to facilitate the efficient migration of the seed toward the galactic center \citep[see e.g][]{Alexander2014,Biernacki2017}.

The studies discussed above suggest that off-center formation of light seeds, followed by a wandering phase before potentially settling in the galactic nucleus, represents a likely evolutionary pathway. However, to date, no systematic study has investigated how this process influences the evolution of the MBH and MBHBs populations across different cosmological times. In this study, we aim to advance in this respect by applying the \LGalaxies{} semi-analytical model (SAM) to merger trees from the \texttt{Millennium} simulation suite \citep{Henriques2015}. 
The paper is organised as follows: Section~\ref{sec:SAM_DESCRIPTIONS} describes the main characteristics of the \LGalaxies semi-analytical model (SAM) and the merger trees from the \texttt{Millennium} simulation suite. In Section~\ref{sec:WanderingModel} we introduce the model included within the SAM to address the off-center formation and wandering phase of the first-forming MBH seeds. In Section~\ref{sec:Results_Nulcear_MBHs} we present how the off-center seed formation and its wandering phase affect the global population of MBHs. Section~\ref{sec:Result_MBHBs} discusses the main effects in the population of MBHBs. Finally, in Section~\ref{sec:Conclusions} we summarize the key  findings. A Lambda Cold Dark Matter $(\Lambda$CDM) cosmology with parameters $\Omega_{\rm m} \,{=}\,0.315$, $\Omega_{\rm \Lambda}\,{=}\,0.685$, $\Omega_{\rm b}\,{=}\,0.045$, $\sigma_{8}\,{=}\,0.9$ and $\rm H_0\,{=}\,67.3\, \rm km\,s^{-1}\,Mpc^{-1}$ is adopted throughout the paper \citep{PlanckCollaboration2014}.

\section{\LGalaxies{} semi-analytical model} \label{sec:SAM_DESCRIPTIONS}

In this section, we present the \LGalaxies{} semi-analytical model (SAM, Bonoli et al in prep). In brief, \LGalaxies{} is based on \LGalaxiesHenriques{} SAM presented in \cite{Henriques2015}, which is designed to track the cosmological assembly of galaxies through a set of analytical equations solved along the merger trees of dark matter halos. The most recent updates to the \LGalaxiesHenriques{} model introduced in \citet{IzquierdoVillalba2020, IzquierdoVillalba2021} and \citet{Spinoso2022} have led to the development of the \LGalaxies{} version. This extended framework allows for the detailed tracking of the formation and evolution of both single and binary MBHs.


\subsection{Dark matter merger trees: N-body simulations}
As \LGalaxiesHenriques{}, \LGalaxies{} is designed to run on top of different dark matter (DM) merger trees extracted from N-body DM-only simulations. Particularly, it has been extensively tested in the \texttt{Millennium} and \texttt{TNG-DARK} suite of simulations \citep[see][]{Henriques2015,Ayromlou2021}. Here, we employ the merger trees extracted from the \texttt{Millennium} (MS, \citealt{Springel2005}) and \texttt{Millennium-II} (MSII, \citealt{Boylan-Kolchin2009}) simulations. MS follows the evolution of $2160^3$ DM particles of mass $8.6\,{\times}\, 10^8\, \mathrm {M_{\odot}}/h$ inside a periodic box of 500 ${\rm Mpc}/h$ on a side, from $z\,{=}\,127$ to the present. MSII tracks the same number of particles but with a mass resolution 125 times higher ($6.885\,{\times}\,10^6\,\mathrm{M_{\odot}}/h$) in a box 125 times smaller ($\mathrm{100\,Mpc}/h$). MS and MSII were stored at 63 and 68 epochs or snapshots with a time separation of $\sim$300 Myr. Structures formed in these simulations were identified by applying friend-of-friend and \texttt{SUBFIND} algorithms and arranged in \textit{merger trees} structures by using the \texttt{L-HALOTREE} code \citep{Springel2001}. Finally, the two simulations were rescaled with the procedure of \cite{AnguloandWhite2010} to match the cosmological parameters provided by \cite{PlanckCollaboration2014}.

\subsubsection{The \textit{Grafting} procedure: Extending the resolution of the underlying dark matter merger trees} \label{sec:EMT}

One of the main limitations of SAMs concerns the minimum halo mass of the merger trees used, i.e. the mass resolution. 
In particular, this limitation results in an incomplete simulated galaxy merger history derived from the underlying dark matter evolution and sets a clear lower limit on the galaxy and MBH evolutionary pathways that can be traced. Although these drawbacks are well known, they are unavoidable given the current difficulties in simulating large cosmological volumes at high resolution. To mitigate these flaws, in this work we use the so-called \textit{grafting} methodology, a two-step process presented in \cite{Angulo2014} and Bonoli et al (in prep). In the first step, one extends the branches of the targeted merger trees below their mass resolution limit using the input of higher-resolution merger trees. This enables the reconstruction of a realistic number of mergers that would otherwise be missed because of mass resolution limits. The second step aims to recover the evolution of structures that are not directly resolved, so that newly initialised galaxies are modelled as evolved systems rather than just pristine reservoirs of baryonic matter.  To do so, each newly resolved halo of the targeted merger tree is initialized with a random galaxy (hosted at the same redshift and halo mass bins) extracted from a run of \LGalaxies{} on top of the merger trees with higher resolution. Following Bonoli et al (in prep), we apply the \textit{grafting} methodology when running \LGalaxies{} on the \texttt{Millennium} simulation merger trees, using  \texttt{Millennium-II} to supply the higher resolution information (see Bonoli et al (in prep) for more details). This procedure enables us to use the results of the \texttt{Millennium} simulation to sample large volumes (crucial for exploring the high-$z$ Universe) while exploiting merger trees as complete as those of the \texttt{Millennium-II} simulation. Notice that we will also make use of the \texttt{Millennium-II} simulation to explore the off-center formation and the subsequent wandering phase of MBH seeds. However, we will not further extend its merger trees due to our absence of higher-resolution merger tree data.



\subsection{From gas to stars: The formation of galaxies}
The galaxy formation model included in \LGalaxies{} is based on the formalism developed by \cite{WhiteandRees1978} and \cite{WhiteFrenk1991} formalism. When a DM halo collapses, a fraction of baryons is captured within it, forming a hot gaseous atmosphere. Over time, this gas cools radiatively and settles at the center of the halo, forming a disc-like structure due to the conservation of angular momentum. The continuous inflow of cold gas triggers star formation (SF), leading to the formation of a stellar disc structure and triggering supernovae explosions (SNe) following the death of massive short-lived stars. This process regulates the assembly of galaxies via the warm-up and ejection of the galaxy cold gas component. The continuous gas accretion of the galaxy central MBH from the hot gas atmosphere also regulates the gas cooling by injecting energy into the intracluster medium. The galaxy can develop a central ellipsoidal component via disc instabilities (secular evolution) or galaxy mergers. The latter process occurs after the merger of the two parent halos on a timescale given by \cite{BinneyTremaine2008}. Major mergers involve two similar galaxies and lead to the formation of a spheroidal galaxy. Conversely, minor mergers involve galaxies with very different masses and allow the survival of the most massive galaxy disc component and the growth of its bulge as a consequence of the integration of the entire stellar mass of the satellite galaxy. 
Environmental processes such as ram pressure stripping or galaxy tidal disruption are also included in the model and have a key role in shaping the galaxy population \citep[see][for further details]{Henriques2015,Henriques2020}. Finally, we stress that, to improve the coarse time resolution offered by the outputs of MS/MSII, \LGalaxies{} performs an internal time interpolation ($\rm {\sim}\,5\,{-}\,20\, Myr$) between snapshots.


\subsection{Massive black holes}
Here, we briefly outline the physical prescriptions implemented in \LGalaxies{} to model the formation and growth of massive black holes, as well as the dynamical evolution of MBHBs.

\subsubsection{A multi-flavour formation of the first MBHs} \label{sec:FormationSMBHs}

The seeding scenario implemented in \LGalaxies{} is presented in Spinoso et al. (in prep) and represents an extension of the one described in \cite{Spinoso2022}, which accounts for a multi-flavour genesis of MBH-seeds. Specifically, the model includes the simultaneous formation of four different categories of MBH seeds: \textit{PopIII remnants}, \textit{runaway stellar mergers} (RSM) seeds, \textit{direct collapse} black-hole seeds (DCBHs) and \textit{merger induced} direct collapse seeds (miDCBH). To provide a physical basis to the DCBH and RSM-seeding scenario, \cite{Spinoso2022} modeled the diffusion and propagation of metals and Lyman-Werner (LW) photons following SF and the associated SNe events. To this end, the authors accounted for both the presence of uniform radiation backgrounds and local variations of the IGM chemical enrichment and LW illumination. These spatial variations were produced around galaxies by intense SF events, which both acted as luminous LW sources and powered the ejection of strong metallic winds thanks to recent SNe explosions. We stress that this MBH seeding model is only applied to the MSII merger trees, whose DM mass resolution allows tracking the physical processes occurring in atomic-cooling halos ($\rm M_{halo}\,{>}\,10^7\, \msun$ at $z\,{>}\,7$), where the seeding of RSM and DCBHs is thought to occur \citep[][]{inayoshi_visbal_haiman2020}. The main differences between the models presented in Spinoso et al. (in prep) and in \cite{Spinoso2022} concern the different treatment of PopIII remnants. In the following, we briefly describe the seeding channels that co-exist in the current version of \LGalaxies{}.\\ 

- \textit{PopIII remnant seeds}:  Although the  MSII simulation can resolve halos of ${\sim}\,10^8\,\msun$, its mass resolution limits  the ability to trace the evolution of the so-called mini-halos ($\rm M_{halo}\,{\sim}\,10^6\,\msun$ at $z\,{>}\,10$) where the formation of ${<}\,10^3\,\msun$ MBH seeds is thought to take place \citep{bromm2013}. To overcome this shortcoming, we follow the sub-grid phenomenological model proposed by Spinoso et al. (in prep), according to which the properties of PopIII remnant seeds are determined from the gas mass available in DM halos as soon as they are newly-resolved in the DM simulation at redshift $z_R$. Thus, every time a halo is newly resolved in the merger tree, it is assigned a probability of being seeded with a light seed according to \citep{Spinoso2022,IzquierdoVillalba2023}: 
\begin{equation} \label{eq:ProbabilitySeeding}
     \mathcal{P}\,{=}\, \min\left[1,\, \mathcal{A} (1+\mathit{z_R})^{\gamma} \left( \frac{\rm M_{halo}}{\rm  \, M_{halo}^{th}}\right)\right]
\end{equation}
\noindent where $\mathcal{A}\,{=}\,0.02$, $\gamma\,{=}\,7/2$ and $\rm M_{halo}^{th}\,{=}\,7\,{\times}\,10^{10}\, \msun{}$ are free parameters chosen in such a way that the occupation fraction of MBHs is similar to the one presented in \cite{Spinoso2022}, obtained taking into account the PopIII population derived form the \texttt{GQd} Press-Schechter SAM \citep{Valiante2021}. Note that we saturate $\mathcal{P}$ to 1 for all the $\rm M_{halo}$ that would provide $\mathcal{P}\,{>}\,1$. 
The end of the PopIII seeding era is set by fixing the values of $\mathcal{P}$ to 0 once the average IGM metallicity in the MSII volume exceeds the critical metallicity threshold $\rm Z_{crit}\,{=}\,10^{-3.5}\,Z_{\odot}$, which corresponds to $z\,{\sim}\,7$ \citep[see Fig.1 in][]{Spinoso2022}. Once assigned the value of $\mathcal{P}$ to the halo, we draw a random value $\mathcal{R} \,{\in}\,[0\,{-}\,1]$ such that if $\mathcal{R}\,{>}\, \mathcal{P}$ a remnant PopIII MBH is seeded. The mass, $\rm M_{BH}(\mathit{z}_R)$, is determined as:
\begin{equation}\label{eq:UnresolvedGrwoth}
    \rm M_{BH}(\mathit{z}_R)\,{=}\,M_{BH}(\mathit{z_f})\,\cdot\,exp\left[\, \mathit{f}_{Edd}^{\,\rm Unres} \frac{1-\eta(\mathit{t})}{\epsilon(\mathit{t})}\frac{\delta t}{\mathit{t}_{Edd}} \right]\,,
\end{equation}
where $\rm M_{BH}(\mathit{z_f})$ corresponds to the mass of the MBH \textit{at the formation redshift}, $z_f$. The latter is chosen randomly between $z_R$ and $z\,{=}\,35$, corresponding to the maximum redshift expected for the formation of PopIII stars \citep[][]{bromm2013}. As mentioned above, we use the gas content of the newly-resolved DM halo at $\mathit{z_R}$ to estimate the properties of PopIII remnant seeds at $z_f$. In particular, we obtain $\rm M_{*}^{Unres}$, i.e. the stellar mass formed during the DM halo unresolved evolution. As in Spinoso et al. (in prep.), $\rm M_{*}^{Unres}$ is derived from a star-formation prescription adapted from Eq.~S14 of \cite{Henriques2015}, which links the production of $\rm M_{*}^{Unres}$ to the cold fraction of the DM halos gas content. 
Although we compute it at $\mathit{z_R}$, we further assume that $\rm M_{*}^{Unres}$ formed at $z_f$ in a single, initial star-formation episode and did not evolve down to $z_R$. This allows us to derive $\rm M_{BH}(\mathit{z_f})$ by randomly sampling a Larson initial mass function \citep[IMF, see][]{larson1998} until $\rm M_{*}^{Unres}$ is reached. 
The actual $\rm M_{BH}(\mathit{z_f})$ is set to the maximum sampled mass, after considering appropriate\footnote{Stellar-evolution models predict that only stars with mass $\rm M_{PopIII}\!>\!25M_\odot$ can produce a PopIII remnant, provided that their mass lies outside the interval $\rm 140\!<\!M_{PopIII}/M_\odot\!<\!260$, where pair-instability leads to the complete destruction of stars at the end of their life \citep[e.g.][]{heger_woosley2002}.} mass-intervals when randomly sampling the Larson IMF. The second term in Eq.~\eqref{eq:UnresolvedGrwoth} 
captures the \textit{unresolved-growth} during the time $\delta t$ elapsed between $\mathit{z_f}$ and $\mathit{z_R}$. The variable $\eta$ ($\epsilon$) is the accretion (radiative) efficiency
, $\rm \mathit{t}_{Edd}\,{=}\,0.45\, \rm Gyr$, and $f_{\rm Edd}^{\rm Unres}\,{=}\,\rm L_{bol}/L_{Edd}$, defined as the ratio between the bolometric ($\rm L_{bol}$) and the Eddington luminosity ($\rm L_{Edd}$). This means that $f_{\rm Edd}^{\rm Unres}$ can be used to parametrize the unknown growth regime of PopIII seeds during their unresolved evolution. In this paper, for simplicity, we assume 
$f_{\rm Edd}^{\rm Unres} \,{=}\,1$ with a fixed spin value, $a$, randomly selected between [0,1)\footnote{The typical formation redshift of the PopIII MBH seeds in our model is approximately $z_f\,{\sim}\,23$ ($6.3\,{<}\,z_f\,{<}\,35$). The usual growth factor is ${\sim}\,5\,{\times}\,\mathrm{M_{BH}}(z_f)$, with a maximum factor found of ${\sim}\,30\,{\times}\,\mathrm{M_{BH}}(z_f)$.}. We stress that the unresolved growth of PopIII remnant seeds will be changed when the off-center seed formation is included (see Section~\ref{sec:WanderingModel}).\\ 
   
 - \textit{RSM seeds}: The formation of intermediate-mass seeds via the collapse of dense, nuclear stellar clusters is also accounted for in the SAM. To this end, \cite{Spinoso2022} adapted the analytical prescription of \citep{DevecchiVolonteri2009} to \LGalaxies{}  by assuming that newly-resolved halos exposed to moderate local LW flux ($\rm {>}\,1\,J_{21}$, where $\rm J_{21}\,{=}\,10^{-21}erg\,cm^{-2}\,s^{-1}\,H_z^{-1}\,sr^{-1}$) and subject to mild chemical enrichment ($\rm {<}\,10^{-3}\,Z_\odot$) can effectively host their first episodes of star formation within their very dense central regions. This can subsequently produce a compact and massive ($\rm {\sim}\,10^{4-6}M_\odot$) nuclear stellar cluster (NSC) which further collapses to a single, massive object of $\rm M_{seed}\,{\sim}\,10^{3-4}M_\odot$ via runaway stellar mergers. This process is modeled closely following \cite{DevecchiVolonteri2009}, and relies on dark matter halo properties, such as halo spin and virial mass, as input parameters \citep[see][]{Spinoso2022}.\\

- \textit{DCBH seeds}: The model assumes that massive seeds of $\rm 10^5\, {\msun}$ can form in halos that: (i) never experienced a SF event, (ii) are illuminated by a high local LW flux ($\rm {>}\,10\,J_{21}$) which can hinder any un-resolved $\rm H_2$-cooling, gas fragmentation and SF events \citep[][]{abel2000,Yoshida2003,shang_bryan_haiman2010,visbal_haiman_bryan2014a,luo2020} and (iii) contain enough chemically-pristine gas ($\rm {<}\,10^{-4}\,Z_{\odot}$). We refer to \cite{Spinoso2022} for further details about the DCBH formation model and the justification of the local LW flux threshold.\\ 
    
- \textit{miDCBHs seeds}: The formation of massive DCBHs can also be triggered during  major mergers of high-$z$ galaxies \citep[][]{MayerAndBonoli2019}. \LGalaxies{} models this formation channel following  \cite{Bonoli2014} which assumes that miDCBHs are formed if: (i) the merger baryonic mass ratio is ${>}\,0.3$, (ii) the merger remnant has a halo mass $\rm {>}\,10^9 M_{\odot}$, (iii) the total cold gas mass carried by the two interacting galaxies ${>}\,8\,{\times}\,10^4\, \msun$, (iv) the two interacting galaxies are disc-dominated (i.e, bulge-to-total ratio ${<}\,0.2$) and, (v) the merger remnant does not already host a nuclear MBH with mass ${>}\,5\,{\times}\,10^4\, \msun$. If all these requirements are met, a $8\,{\times}\,10^4\,{\msun}$ seed is formed at the centre of the galaxy.\\  

\begin{figure}    
\includegraphics[width=1\columnwidth]{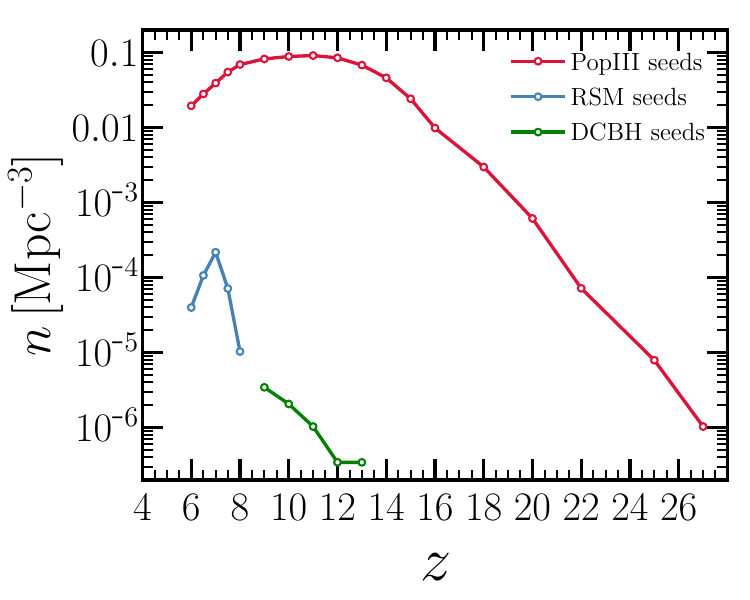}
\caption[]{Redshift evolution of the number density of newly formed seeds predicted by \LGalaxies{} when run on top of the \texttt{Millennium-II} merger trees. Red, blue and green lines correspond to the PopIII remnants, RSM and DCBH seeds, respectively.}
\label{fig:NewlyFormedSeeds}
\end{figure}

To guide the reader, Fig.~\ref{fig:NewlyFormedSeeds} shows the redshift evolution of the number density of newly formed seeds predicted by \LGalaxies{}. As observed, the number density of these objects is primarily dominated by PopIII remnants, reaching values as high as $\rm {\sim}\,0.1\, Mpc^{-3}$ at $z\,{\sim}\,10$. The second most numerous population consists of RSM seeds, which begin forming at $z\,{\sim}\,8$ and peak at $z\,{\sim}\,7$, with a maximum number density of ${\sim}\,10^{-4}\,\rm  Mpc^{-3}$. Conversely, DCBHs are the least dominant population 
Their formation starts at $z \,{\sim}\,12$ and ends at $z\,{\sim}\,8$, moment at which they reach the highest formation rate with a number density of $\rm 10^{-5}\, Mpc^{-3}$. For simplicity, the figure does not include the seeds formed via the miDCBH channel, which occurs at $z\,{\lesssim}\,4$ and feature number densities $\rm {\lesssim}\,10^{-5}\, Mpc^{-3}$. Based on these results, the population of MBHs formed within \LGalaxies{} is overwhelmingly dominated by PopIII remnants. As we will see later, this will have a significant impact on the formation of the MBH population, since PopIII remnant seeds are expected to form off-centre, and their relatively light mass will inevitably cause them to drift through the galaxy, far from the galactic nucleus.

\subsubsection{The growth of massive black holes}

\LGalaxies{} assigns to newly formed MBH a random spin ($\chi$) whose modulus is evolved consistently due to accretion of gas and MBH mergers \citep{IzquierdoVillalba2020}. The growth of MBHs is mainly governed by cold gas accretion after galaxy mergers of disc instabilities. These are assumed to drive a fraction of the galaxy cold gas component towards the galactic center. This gas is assumed to form a gas reservoir around the MBH ($\rm M_{res}$) which is progressively accreted in two phases \citep{IzquierdoVillalba2020,IzquierdoVillalba2024}. The first one is either Eddington-limited or Super-Eddington, depending on the environment in which the MBH is embedded. For large gas reservoirs around the MBH ($\mathcal{R} \,{=}\, \rm M_{res}/M_{BH} \,{>}\,2\,{\times}10^4$) and large gas inflow after a galactic merger or disc instability ($\rm \dot{M}_{inflow} \,{=}\, \Delta \rm M_{BH}^{gas} / t_{\rm dyn}^{Gal} \,{>}\,10\, M_{\odot}/yr$) the MBH triggers a \textit{Super-Eddington} event, characterized by an Eddington-ratio ($f_{\rm Edd}\,{=}\,\rm L_{bol}/L_{Edd}$) of:
\begin{equation}
    f_{\rm Edd} = B(\chi) \left[\frac{0.985}{{\rm \dot{M}_{Edd}/\dot{M}}+C(\chi)}+\frac{0.015}{{\rm \dot{M}_{Edd}/\dot{M}}+D(\chi)}\right],
\end{equation}
where $\rm \dot{M}$ ($\rm \dot{M}_{Edd}$) is the (Eddington) MBH accretion rate and $\rm L_{bol}$ ($\rm L_{Edd}$) the MBH (Eddington) luminosity. The variables $B(\chi)$, $C(\chi)$, $D(\chi)$ correspond to functions taken from \cite{Madau2014}. To guide the reader, the typical Eddington ratios ($\rm \dot{M}  /\dot{M}_{Edd}$ ratios) predicted in our model vary between $1\,{-}\,4$ ($10\,{-}\,1000$). 
If the super-Eddington conditions are not met, an \textit{Eddington} limit growth is assumed, whose $f_{\rm Edd} \,{=}\, 1$. On the other hand, the second phase starts when the MBH consumes a fraction of the initial gas mass reservoir, and it is set to:
\begin{equation}
    f_{\rm Edd} \,{=}\, \left[1 + \left((t - t_0)/t_Q\right)^{1/2}\right]^{-2/\beta},
\end{equation}
where $ t_Q \,{=}\, t_d\,\xi^{\beta}/(\beta \ln 10)$ is the time-scale at which $f_{\rm Edd}$ declines. $t_d \,{=}\, 1.26{\times}10^8 \, \rm yr $, $\beta \,{=}\, 0.4$ and $\xi \,{=}\, 0.3$. The value of these variables are based on \cite{Hopkins2009} which indicated that models of \textit{self-regulated} MBH growth require $0.3\,{<}\,\beta\,{<}\,0.8$ and $0.2\,{<}\,{\xi}\,{<}\,0.4$. All the free parameters involved in the MBH growth are calibrated to reproduce the local galaxy-MBH correlation and the stochastic gravitational wave background (sGWB) reported by Pulsar Timing Arrays \citep[PTAs,][]{Antoniadis2023,Agazie2023,Reardon2023}.

\subsection{Massive black hole binaries} \label{sec:MBHB_theory}

Rather than assuming that massive black holes (MBHs) instantly sink to and merge at the galaxy nucleus following a galaxy-galaxy merger, \LGalaxies{} models the entire lifetime of massive black hole binaries \citep[MBHBs,][]{IzquierdoVillalba2021}. The evolution of MBHBs is divided into three different phases \citep{Begelman1980}: \textit{pairing}, \textit{hardening} and \textit{gravitational wave phase}. The first one starts after the merger of two galaxies. The time spent by the MBH in this phase is computed using the expression of \cite{BinneyTremaine2008}, assuming that the initial separation between the two MBHs corresponds to the radius at which the tidal forces removed 80\% of the satellite stellar mass. The second and third phases begin when the satellite MBH reaches the galaxy nucleus, and it binds to the central MBH thus forming a MBHB. The initial binary eccentricity is randomly drawn in the range $[0-0.99]$ and the initial semi-major axis is set to the distance at which the stellar content of the galaxy (distributed according to a Sérsic model) equals the mass of the secondary (i.e lightest) MBH. The eccentricity and separation of the MHBH are then evolved self-consistently depending on whether the environment in which the MBH resides is stellar or gas dominated \citep{Sesana2015,Dotti2015}. For multiple galaxy mergers, triple MBH interactions can occur whose outcome is modeled according to \cite{Bonetti2018ModelTriplets}. Finally, \LGalaxies{} also takes into account the growth of MBHs and MBHBs during the pairing and hardening phases \cite[see][for further details]{IzquierdoVillalba2021}. 

\section{The off-centered formation and wandering phase of the first MBH seeds} \label{sec:WanderingModel}

Recent hydrodynamical simulations have shown that light MBH seeds rarely form and reside at the center of their high-$z$ hosts. Instead, the lack of a deep potential and the irregular morphologies of these galaxies make MBH seeds undergo a random walk in and out of the galactic centre \citep{Smith2018,Regan2020,Pfister2019,Bellovary2019,Ma2021}. So far, galaxy formation models have generally neglected this \textit{off-centered} seed formation and the subsequent \textit{wandering phase}. Here, we present all the assumptions made to model this feature (hereafter \Wandering{} model). Precisely, we adapt the model to assess the site at which the first seeds appear and
account for the time required for them to sink to the galactic centre, along with any potential growth phases during this period.




\subsection{The formation place of MBH seeds} \label{sec:Formation_Place}

Determining the initial position at which MBH seeds form ($r_0$) is crucial since it determines the time spent by the MBH seed in a wandering phase before reaching the galactic nucleus. In our model, we assume that DCBH, miDCBH and RSM form at the galactic centre and their large masses (including the NSC mass for the RMS) ensure their stability at their host centres \citep[see e.g][]{Alexander2014,Biernacki2017,Pfister2019}. In contrast, for PopIII remnants, we follow the findings of \cite{Shi2015} and \cite{Smith2018}, assuming they form off-centered. Specifically, the high-resolution hydrodynamical simulations by \cite{Shi2015} showed that by $z \,{\lesssim}\, 15$, half of the PopIII remnants are located within 10\% of the host halo virial radius ($\rm R_{vir}$). Furthermore, their radial distribution exhibits a clear peak at ${\sim}\,0.05\,\rm R_{vir}$, with a tail extending out to the full virial radius. To mimic these results, as soon as a PopIII remnant is formed in a DM halo its position is randomly chosen according to a log-normal distribution with median $0.1\,\rm R_{vir}$ and variance $0.85\,\rm R_{vir}$. 
Besides that, to develop an off-center PopIII formation model that aligns more closely with recent hydrodynamical simulation results, we also consider the possibility that multiple PopIII remnants can form off-center within a single, newly-resolved halo \citep[][]{Smith2018}. To do so, we use the modifications of multi-PopIII seed formation presented in Spinoso et al. (in prep). Specifically, it extends the PopIII-remnant formation scenario described in Sect.~\ref{sec:FormationSMBHs} by allowing that a number $\rm N_{seeds}\,{=}\,M^{Cold}_{gas}/M_J\,{\geq}\,1$ of PopIII remnants can form, where $\rm M^{Cold}_{gas}$ and $\rm M_J$ are the cold-gas content and Jeans mass of the halo hosting the formation of PopIII remnants, respectively. We stress that the multi-PopIII formation described here is only part of the \Wandering{} model and not applied in the \Fiducial{} one. To guide the reader, the distribution of the formation places of light seeds can be found in Appendix~\ref{Appendix:R0_and_Twand}. According to the model, ${>}\,50\%$ of light seeds form at distances between 100 pc to 500 pc from the halo (or galaxy) centre.

\begin{figure*}    
\centering  
\includegraphics[width=2\columnwidth]{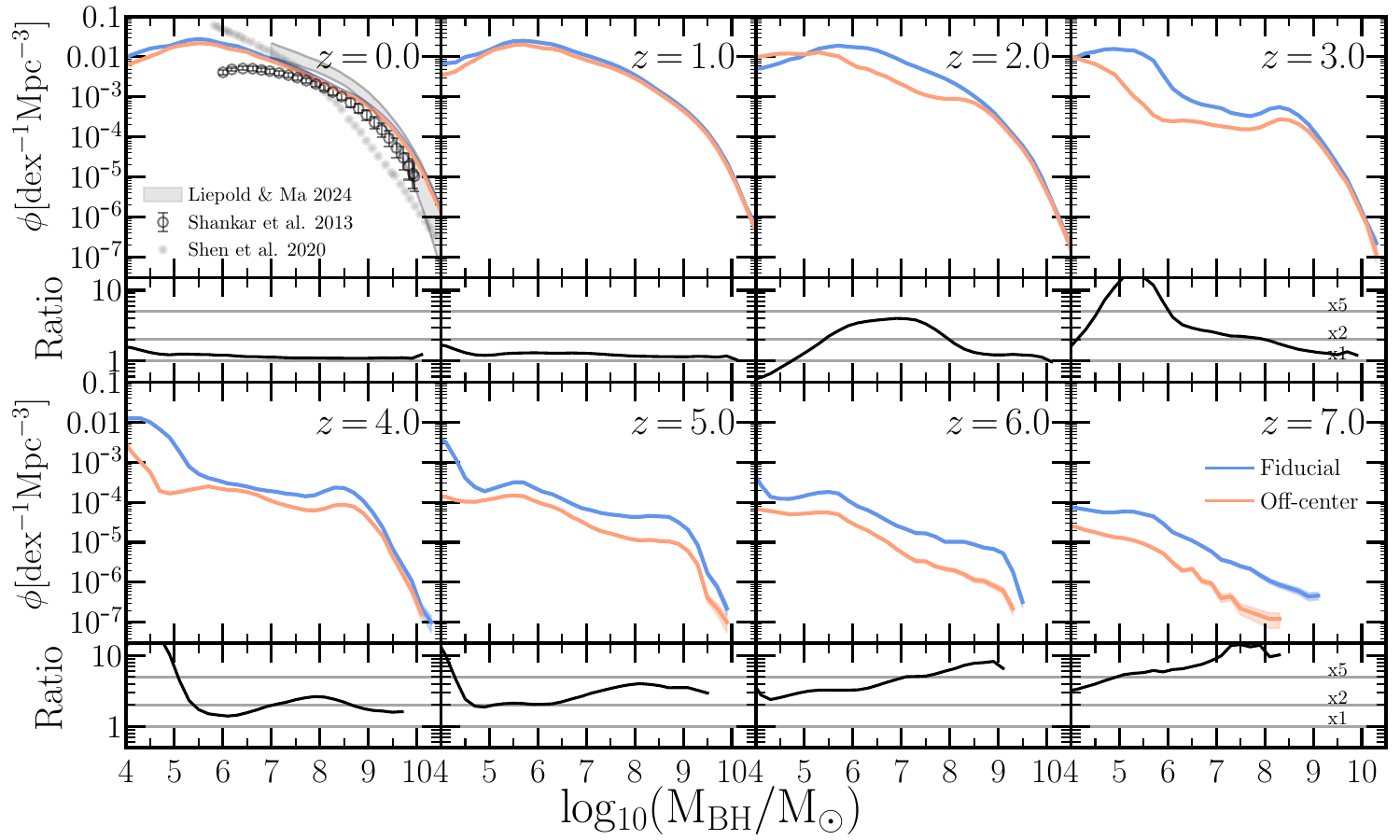}
\caption[]{\textbf{Upper panels}: Black hole mass function (BHMF) at $z\,{=}\,0,1,2,3,4,5, 6$ and $7$. The results correspond to the \LGalaxies{} SAM run on the \texttt{Millennium} merger trees after applying the \textit{grafting} methodology presented in {\protect Section~\ref{sec:EMT}}. The $z\,{=}\,0$ observations correspond to {\protect \cite{Shankar2013}, \cite{Shen2020} and \cite{LiepoldAndMa2024}}. Blue and orange lines represent the \Fiducial{} and \Wandering{} model predictions. Shaded areas correspond to the Poisson error. \textbf{Lower panels}: Ratio between the predictions of the \Wandering{} and \Fiducial{} models. The horizontal gray lines are placed at $1,2, 5$.} 
\label{fig:BlackHoleMassFunction}
\end{figure*}

\subsection{The duration of the off-centered phase} \label{sec:WanderingTime}

To determine the duration of the wandering phase of an MBH seed ($t_{\rm w}^{\rm BH}$) we make a very conservative assumption in which the MBH seed is moving inside a spherical and singular isothermal galaxy sphere. Under these conditions, the dynamical friction timescale, $t_w$, can be written as \citep{BinneyTremaine2008}:
\begin{equation}  \label{eq:DynamicalFriction}
    t_{\rm w}^{\rm BH} \,{=} \, 19 \, f_s \, f(\varepsilon)  \left( \frac{r_0}{5 \, \rm kpc} \right)^2 \left( \frac{\sigma}{200 \,\rm km/s}\right) \left( \frac{10^8 \, \rm M_{ \odot}}{\rm M_{BH}} \right) \, \frac{1}{\Lambda}\, \rm [Gyr] , 
\end{equation}
where $r_0$ is the distance of seed formation from the halo centre (see Section~\ref{sec:Formation_Place}), $f(\varepsilon)\,{=}\,\varepsilon^{0.78}$ is a function that depends on the circularity $\varepsilon$ of the MBH orbit \citep{Lacey1993}, $\sigma$ is the velocity dispersion of the galaxy, $\rm M_{BH}$ is the mass of the MBH seed 
and $\rm \Lambda\,{=}\,\ln(1 + M_{stellar}/M_{BH})$ is the Coulomb logarithm \citep{MoWhite2010}. For simplicity, we fix $\varepsilon\,{=}\,0.5$. Finally, the variable $f_s$ is included to mimic the stochastic insparalling of MBHs seen in simulations of clumpy (gas-rich) and barred galaxies \cite[see][]{Lupi2015,Tamburello2017,Pfister2019,Bortolas2020,Bortolas2022}. The parameter $f_s$ was introduced in \cite{IzquierdoVillalba2023} and modeled as a log-normal distribution with a median value of 0.2 and a variance of 0.6. The choice of these values was motivated by the shape of the resulting log-normal distribution, which peaks at ${\sim}\,1$ and features a positive skewness. If the MBH seed forms in a gas-rich or barred galaxy, $f_s$ is computed from the above distribution. Otherwise, $f_s$ is set to 1. The distribution of the time spent by the light seeds wandering within the galaxy after formation can be found in Appendix~\ref{Appendix:R0_and_Twand}. According to the model ${>}\,70\%$ of the light seed wanders for more than 1 Gyr before settling at the halo centre. Note that this is longer than the Hubble time for $z\,{>}\,6$.

Finally, we highlight that the shrinking timescale described by Eq.~\eqref{eq:DynamicalFriction} is a simplified approximation, as it assumes that MBHs evolve within a smooth, spherical, isothermal potential. In reality, the structure of galaxies is more complex. For example, studies by \cite{Pfister2019}, \cite{Bellovary2019}, and \cite{Ma2021} have shown that irregularities in the gas and stellar distributions can lead to gravitational scattering of MBHs, significantly delaying or even preventing their inward migration toward the galactic center. In addition to galactic structure, the mass of the black hole seed plays a crucial role. \cite{Ma2021} found that low-mass MBH seeds struggle to sink efficiently in typical high-$z$ galaxies and may require embedding within dense stellar clusters to overcome dynamical friction (see also \citealt{Alexander2014,Biernacki2017}). Furthermore, Eq.~\eqref{eq:DynamicalFriction} does not account for natal kicks that can be imparted to PopIII MBH seeds at formation. As shown by \cite{Whalen2012}, such kicks may eject low-mass seeds from their host halos, severely inhibiting their ability to return and settle in the galactic center.

\subsection{The growth of seeds during the off-center phase}
An important aspect to consider is the growth of MBH seeds during their wandering phase after their off-center formation. Analysis of the hydrodynamical simulations \textit{Renaissance} by \cite{Smith2018} showed that PopIII remnants experience minimal mass increase during their off-center phase, with a maximum growth of only about 10\% relative to their initial seed mass (see also \citealt{Liu2020} for similar results). In this work, we incorporate these findings by assuming that MBH seeds, which form off-centered accrete gas from the galaxy cold gas disc at a fixed Eddington ratio of $f_{\rm Edd} \,{=}\, 10^{-4}$ (see Figure 3 in \citealt{Smith2018}). 
Analogously, during the unresolved growth phase of PopIII remnants described in Section~\ref{sec:FormationSMBHs} (see Eq.\ref{eq:UnresolvedGrwoth}), we set the Eddington ratio $f_{\rm Edd}^{\rm Unres} \,{=}\, 10^{-4}$.

\section{Implications of the off-centred seed formation in the population of nuclear MBHs}  \label{sec:Results_Nulcear_MBHs}

In this section, we investigate how the off-centered formation of MBH seeds impacts the overall MBH population. In particular, we examine the resulting black hole mass and quasar luminosity functions. We then analyse the imprint of off-center formation and the subsequent wandering phase on the scaling relations at both high and low redshifts. A comprehensive description of the trends observed in the \Fiducial{} model will be presented in Bonoli et al. (in prep).

\begin{figure}    
\centering  
\includegraphics[width=1.\columnwidth]{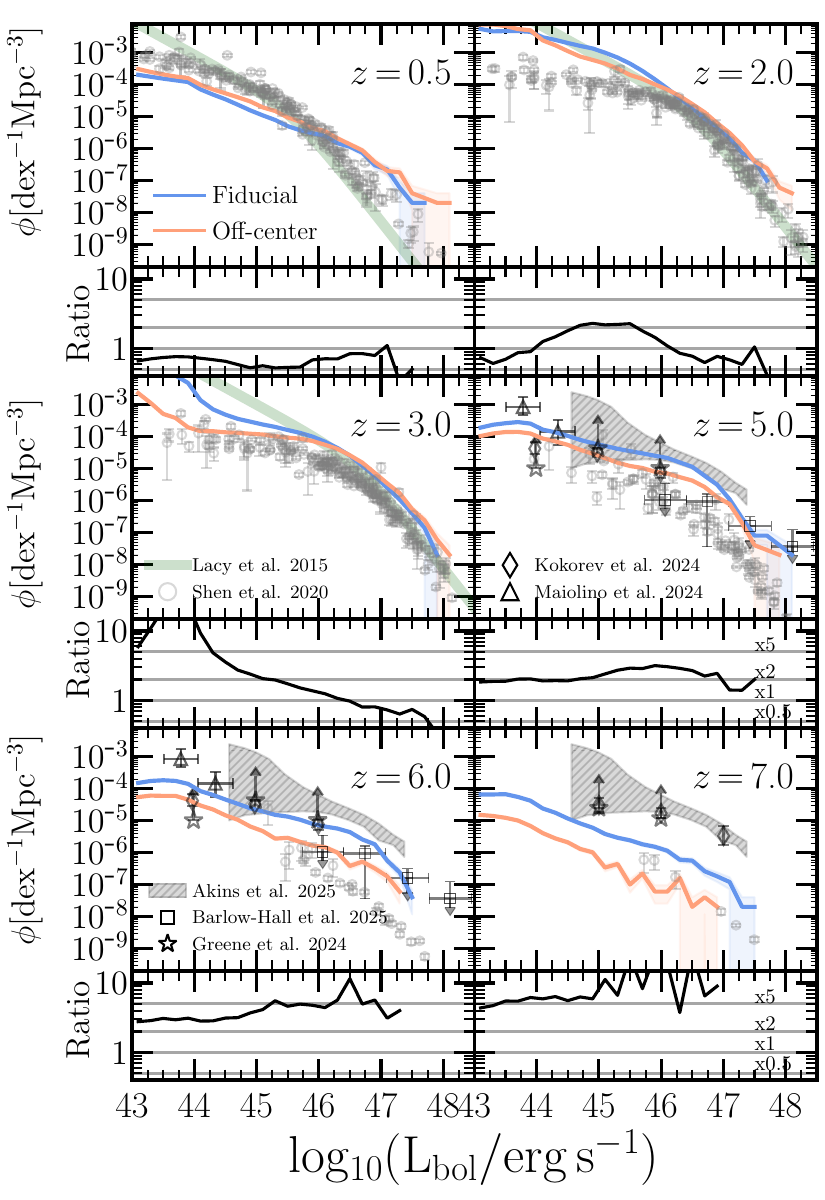}
\caption[]{\textbf{Upper panels}: Quasar luminosity function at $z\,{=}\,0.5,2,3,5,6$ and $7$ predicted by the \Fiducial{} (blue) and \Wandering{} (orange) model. The results of \LGalaxies{} are presented for the \texttt{Millennium} trees with the \textit{grafting} methodology. Shaded areas correspond to the the Poisson error. Observational data corresponds to \protect{\cite{Shen2020,Lacy2015,Kokorev2024,Greene2024,Matthee2024,Akins2024} and \cite{BarlowHall2025}}. \textbf{Lower panels}: Ratio between the predictions of the \Wandering{} and \Fiducial{} models. The horizontal grey lines are placed at $0.5,1,2$ and $5$ to guide the eye.} 
\label{fig:LuminosityFunction}
\end{figure}


\subsection{The assembly of MBHs: The black hole mass and quasar luminosity function} \label{sec:BHMFandLF}

The evolution of the MBH mass functions in the \Fiducial{} and \Wandering{} models is shown in Fig.~\ref{fig:BlackHoleMassFunction}. At $z\,{\sim}\,7$ the \Fiducial{} model features ${>}\,5$ times more MBHs with mass ${>}\,10^5 \, \msun$ than the \Wandering{} case. This behaviour changes towards lower redshifts. At $z\,{\sim}\,5$, the \Wandering{} model reduces the differences with the \Fiducial{} one a factor ${\sim}\,2\,{-}\,3$ for objects with $\rm 10^{5} \,{<}\,M_{BH}\, {<}\,10^{8.5}\, \msun$. 
This trend continues at $3 \,{<}\, z \,{<}\, 5$ for MBHs with $\rm M_{BH}\,{>}\,10^{8.5}\, \msun$ but differences rise again in the range of $10^4\, \msun \, {<}\, M_{BH}\,{<}\,10^6 \, \msun$ where the \Fiducial{} model predicts up to 10 times larger abundances. 
As expected, this behavior is driven by the off-centered seed formation, which delays the assembly of intermediate mass MBHs. By $z\,{\sim}\,2$ the two models show similar abundances at $\rm {>}\,10^{8.5}\, \msun$ and the differences are reduced below a factor 2 for MBHs with $10^4 \, \msun \,{<}M_{BH}\,{<}\,10^{5}\, \msun$. Despite that, the \Wandering{} model still predicts approximately 3 times fewer MBHs at $10^6\, \msun \, {<}\, M_{BH}\,{<}\,10^8 \, \msun$ than the \Fiducial{} one. Finally, at $z\,{\leq}\,1$ both models exhibit similar behavior, showing a good agreement with the observational constraints of \cite{Shankar2013} and \cite{LiepoldAndMa2024}, 
highlighting that off-center seed formation has a minimal impact on the abundance of low-$z$ MBHs. We notice that the only notable difference at $z\,{\leq}\,1$ occurs at masses  ${<}\,10^{6} \, \msun$, where the \Fiducial{} model predicts approximately $1.5$ times more objects than the \Wandering{} one. We stress that the convergence between the two models at low redshift is primarily driven by the fact that MBHs initially formed off-center have had sufficient time to migrate toward the galactic nucleus. This convergence occurs earlier for more massive black holes, as they reside in more massive galaxies whose deeper gravitational potentials and denser stellar environments enable more efficient inward migration. In contrast, low-mass galaxies are less effective at facilitating this process, leading to a delayed convergence at the low-mass end of the MBH mass function. Moreover, the peak of the galaxy merger rate occurs at lower redshifts for less massive systems. As a result, MBHs in low-mass galaxies must typically wait longer for a merger event capable of efficiently triggering their growth, once they reach the galactic nucleus (see, e.g., Fig. 8 in \citealt{Volonteri2020}).\\

In addition to the assembly of MBHs, off-centered seed formation can also influence the observable population of accreting MBHs. To investigate this, Fig.~\ref{fig:LuminosityFunction} shows the evolution of the predicted quasar bolometric luminosity functions (LF). These predictions are compared with observational data from \cite{Shen2020}, which compiles infrared, optical, UV, and X-ray measurements, \cite{BarlowHall2025} which gathers X-ray data, as well as recent JWST results from \cite{Akins2024}, \cite{Greene2024}, \cite{Matthee2024} and \cite{Kokorev2024}. It is important to emphasise that the JWST LFs presented here consist of Little Red Dots samples (LRDs), which imply that not all objects in the sample must necessarily be AGNs\footnote{The sample of \cite{Greene2024} is most likely composed of AGNs as the LRDs are selected by colour compactness and by the presence of broad-line emissions. \cite{Kokorev2024} uses the same color selection as \cite{Greene2024}, but does not check for broad lines. \cite{Akins2024} uses different colour selections than the two previous cases and assumes two extreme cases: the LRDs of the same are all AGN or all galaxies. \cite{Matthee2024} selects the sample by looking for broad line emissions in slitless/grism spectroscopy, implying that is the best approach to obtain an actual AGN luminosity functions.}.  
Similar to the mass function, the most significant differences appear at high-$z$. Specifically, at $6\,{<}\,z\,{<}\,7$ the \Wandering{} model predicts about 5 to 10 times fewer active MBHs than the \Fiducial{} model, regardless of luminosity. This behavior is anticipated, as the \Wandering{} model shows a lower number density of MBHs with ${>}\,10^4, \msun$, as seen in Fig.~\ref{fig:BlackHoleMassFunction}. Interestingly, the \Fiducial{} model aligns better with the observed high-$z$ LF, especially when constrained by the photometrically selected JWST objects \citep{Kokorev2024, Akins2024}. At $z\,{=}\,5$ differences between the two models persist but decrease to about a factor of 2 at luminosities at $\rm 10^{44-45}\, erg/s$ and at luminosities above $10^{47}\, \rm erg/s$. Despite these differences, both models generally agree well with the LFs reported in \cite{Greene2024, Kokorev2024} and \cite{Akins2024}. At $z\,{=}\,3$, both models produce similar results for luminosities ${>}\,10^{46}\, \rm  erg/s$. However, differences become more pronounced at lower luminosities ($10^{43-44} \, \rm erg/s$), with the \Fiducial{} model predicting up to 10 times more objects than the \Wandering{} model. Interestingly, this lower number of low-luminosity MBHs in the \Wandering{} model better reproduces the LF shape observed in \cite{Shen2020}. Similar trends are observed at $z\,{=}\,2$, though the differences at low luminosities nearly disappear. By $z\,{=}\,0.5$, the trend reverses: the \Wandering{} model predicts about 1.5 times more active MBHs than the \Fiducial{} model. This shift results from the delayed assembly of the MBH population in the \Wandering{} model, which causes massive galaxies to quench later. Consequently, MBHs in the \Wandering{} scenario tend to reside in galaxies with larger gas reservoirs, allowing sustained MBH accretion over longer periods after galaxy mergers or secular processes.\\

In summary, off-center MBH seed formation does not hinder the assembly of a MBH population consistent with current observational constraints of low-$z$ (active and inactive) MBHs. 
However, significant differences emerge at high-$z$ ($z\,{>}\,6$), particularly in the quasar LF. Therefore, obtaining a robust observational estimate of the number density of MBHs with bolometric luminosities $10^{44-47}\, \rm erg/s$ at $z\,{>}\,6$  (a range our model indicates could vary by up to a factor of five) will be essential to constrain
the early growth and dynamics of the first MBH seeds. 

\subsection{The galaxy and massive black hole scaling relation}


\begin{figure}    
\centering  
\includegraphics[width=1.\columnwidth]{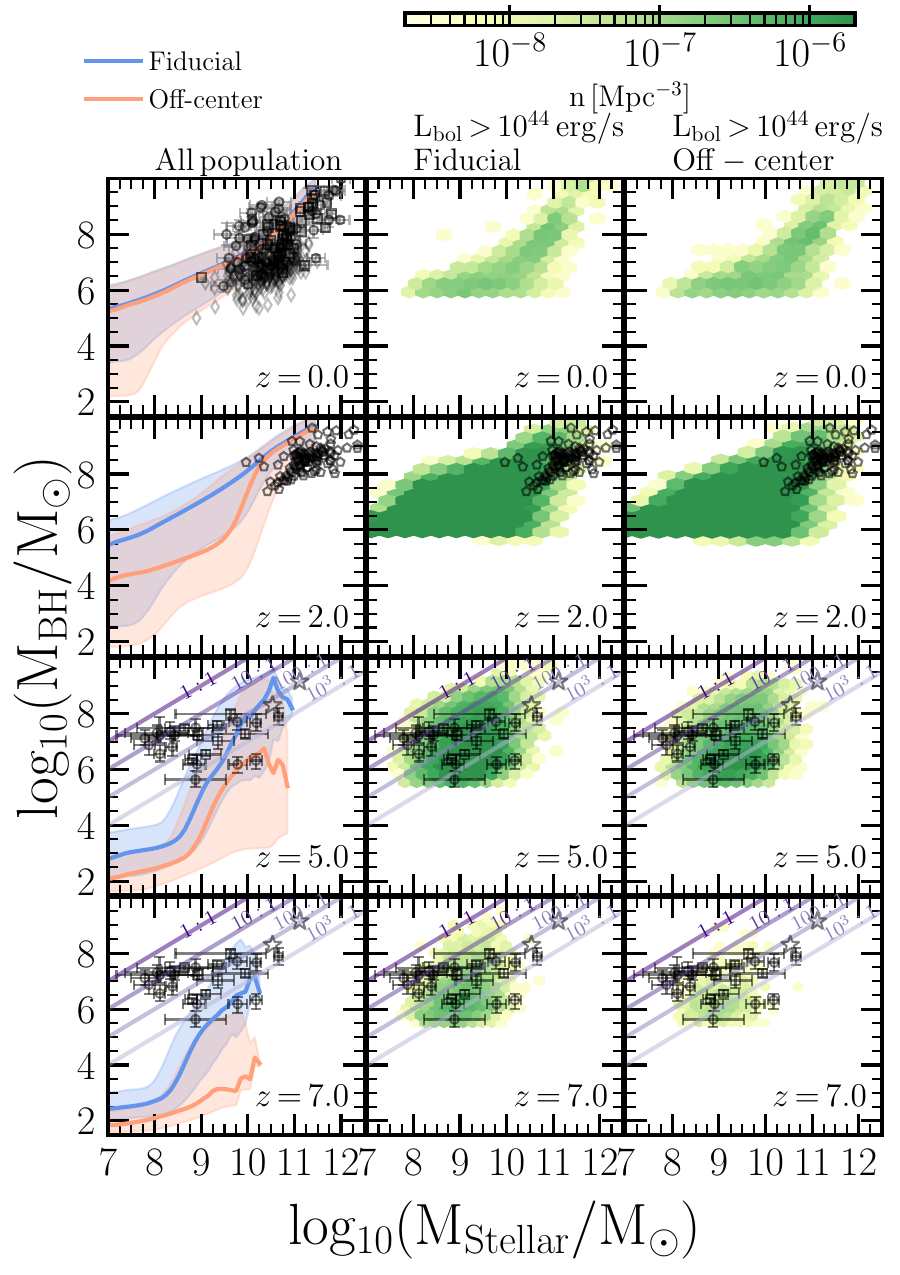}
\caption[]{Scaling relations at predicted by \LGalaxies{} applied in the \texttt{Millennium} trees with the \textit{grafting} methodology. \textbf{Left panel:} Median black hole mass ($\rm M_{BH}$) at fixed stellar mass ($\rm M_{Stellar}$). The shaded areas correspond to the percentile $\rm 16^{th}\,{-}\,84^{th}$. Blue color represents the results for the \Fiducial{} model while orange corresponds to the \Wandering{} one. \textbf{Central and right panels:} $\rm M_{BH} \,{-}\, M_{Stellar}$ plane for MBHs with $\rm L_{bol}\,{>}\,10^{44} erg/s$. As shown, that population far beyond the $\rm 16^{th}\,{-}\,84^{th}$ percentiles. The observational constraints are: $z\,{=}\,0$ {\protect \citealt{Erwin2012} (squares) \citealt{ReinesVolonteri2015}  (diamonds) and \citealt{Capuzzo2017} (circles)}; $z\,{=}\,2$ \cite{Shu2020} (pentagons) and at $z\,{\geq}\,5$ {\protect \citealt{Maiolino2023} (circles), \citealt{Harikane2023} (squares) and \citealt{Ding2023} (stars)}.}
\label{fig:ScalingRelation}
\end{figure}

Studies at low-$z$ have shown a correlation between black hole mass and host galaxy stellar mass. 
This relation is thought to arise from a simultaneous growth, regulated by AGN feedback, of galaxies and nuclear MBHs during galaxy mergers or secular processes \citep[e.g.,][]{DiMatteo2005,Croton2006}. To better understand this relation, predictions from the \Fiducial{} model are presented in  Fig.~\ref{fig:ScalingRelation}. First, we compare these predictions to observational data to establish a baseline. Then, we discuss the differences introduced by the off-center model, which are shown in subsequent analyses. 
At $z\,{\geq}\,5$ the median relation of the \Fiducial{} model falls below the recent constraints derived from JWST data \citep{Maiolino2023, Harikane2023, Ding2023}. Specifically, the typical MBHs in galaxies of $\rm 10^{8}\,{<}\,M_{Stellar}\,{<}\,10^{9}$ are ${\sim}\, 4 \, \rm dex$ smaller ($\rm 10^{2-4} \, \msun$) compared to current observational estimates ($\rm {\sim}\, 10^6 \, \msun{}$). This discrepancy is less pronounced at higher stellar masses ($\rm M_{Stellar}\,{>}\,10^9 \, \msun$), where the model predictions show some agreement with some high-$z$ objects. Since JWST only selects active MBHs (such as broad-line AGNs; see e.g., \citealt{Harikane2023, Maiolino2023, Matthee2024}), the right panel of Fig.~\ref{fig:ScalingRelation} offers a fairer comparison, by showing the scaling relation for galaxies hosting MBHs with bolometric luminosities $\rm L_{bol}\,{>}\,10^{44} \, erg/s$. The population of active MBHs predicted by the \Fiducial{} model is ${>}2\sigma$ away from the relation of the whole MBH population (left panel) and aligns well with high-$z$ AGNs. This suggests that active high-$z$ MBHs occupy a distinct region in the relation (more than $2\sigma$ beyond the median relation) compared to the overall MBH population \citep[see discussion in][for differences in the scaling relation due to biases or intrinsic differences]{Lauer2007,ReinesVolonteri2015}. 
Finally, the \Fiducial{} model predictions at low-$z$ ($z\,{<}\,2$) match well with the constraints provided by \cite{Erwin2012}, \cite{ReinesVolonteri2015} and \cite{Capuzzo2017}. We stress that at $z\,{=}\,2$ there are some differences between the model and the observations. These can be due to the fact that \cite{Shu2020} data corresponds to broad line AGNs, selected according to several specific cuts. In contrast, our median corresponds to the whole MHB population.


Since the correlation between MBHs and galaxies stems from their synchronised growth, it is possible that the off-center periods included in the \Wandering{} model could disrupt or modify the scaling relation observed in the \Fiducial{} case. To investigate this, in the left panel of Fig.~\ref{fig:ScalingRelation} we show in orange the predicted scaling relation from the \Wandering{} model. Overall, the trend is similar to that of the \Fiducial{} case, however, the normalization of the relation is notably affected and exhibits a redshift evolution. At $5\,{<}\,z\,{<}\,7$ the \Wandering{} model predicts smaller MBHs at a given stellar mass. The difference is typically $\rm {\sim}\,0.5\,dex$, but can reach up to $\rm \,{\sim}\,2\, dex$ at $z\,{=}\,7$ for galaxies of $\rm 10^{9}\,{<}\,M_{Stellar}\,{<}\,10^{9.5}\, \msun$. By $z\,{=}\,2$, the correlation found by the \Wandering{} model aligns with the \Fiducial{} one for galaxies with $\rm M_{Stellar}\,{>}\,10^{10}\, \msun$. For smaller systems, the \Wandering{} model predicts lighter MBHs, with differences up to $2\, \rm dex$. This systematic offset disappears by $z\,{=}\,0$, where both models converge. Regarding AGNs, the right panel of Fig.~\ref{fig:ScalingRelation} shows that the predictions of the \Wandering{} model are similar to those of the \Fiducial{} one, remaining consistent with the existence of the overmassive high-$z$ objects reported by \cite{Maiolino2023} and \cite{Harikane2023}. Based on these findings, the model predicts that off-center MBH formation plays a significant role in shaping the correlation between MBHs and galaxies at high redshift ($z\,{>}\,5$). 
Although the number density of the biased population of active MBHs is affected (see Fig.~\ref{fig:LuminosityFunction}), the expected correlation between MBH mass and galaxy mass remains unchanged. Since these MBHs are the only group currently observable, the existing constraints on galaxy-MBH scaling relations at high-$z$ derived from JWST data do not provide strong insights into the early stages of MBH seed formation and their initial dynamics.




\subsection{The mass contribution from gas accretion}

\begin{figure}    
\centering  
\includegraphics[width=1.\columnwidth]{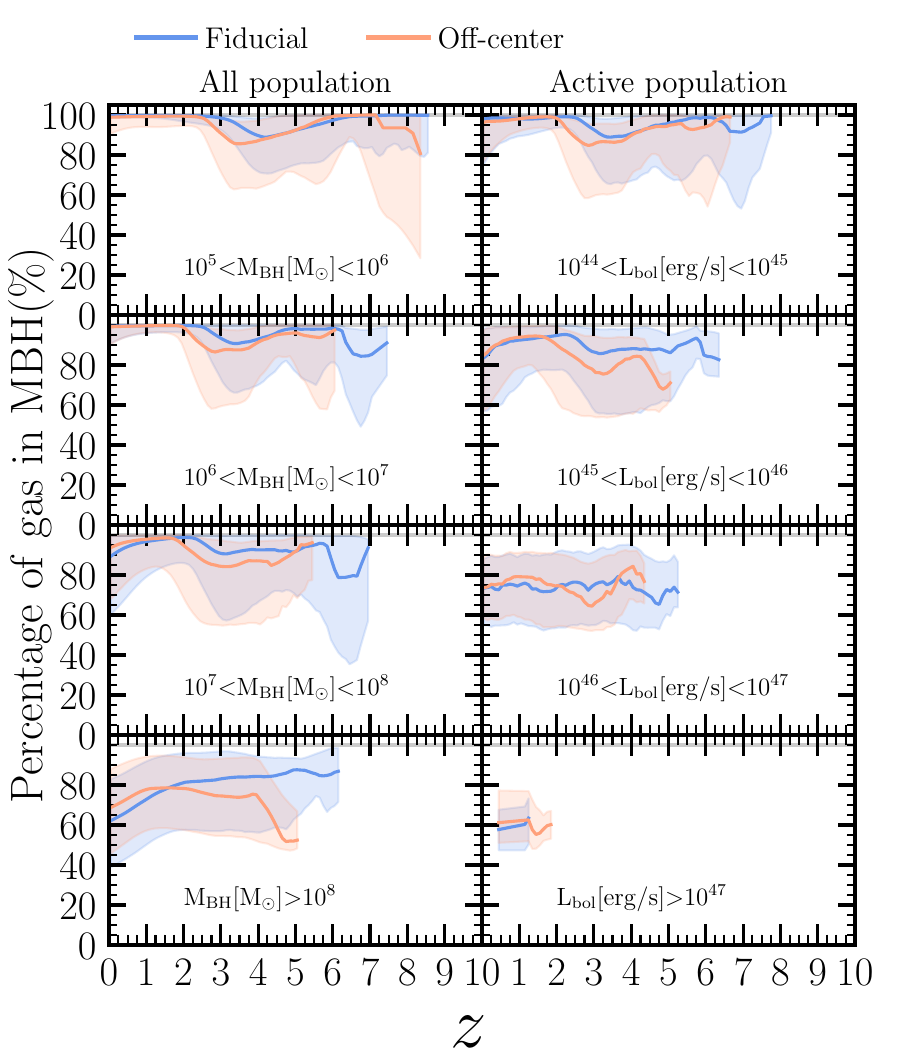}
\caption[]{Fraction of mass acquired by the MBH via gas accretion. While the left panels represent the entire MBH population divided in mass bins, the right panel is restricted to active MBHs, divided in different bolometric luminosities. Orange and blue lines represent the results for \Fiducial{} and \Wandering{} one. Shaded areas correspond to the $\rm 16^{th}\,{-}\,84^{th}$ percentiles of the distribution. 
}
\label{fig:Gas_Merger_accretion}
\end{figure}

The left panel of Fig.~\ref{fig:Gas_Merger_accretion} shows the fraction of mass gained by the MBH through gas accretion, as opposed to the mass gained by mergers. 
As illustrated, there is a general trend that the lower the MBH mass, the larger the proportion of the mass gained via gas accretion. However, differences emerge when comparing the \Fiducial{} and \Wandering{} models. For systems with $\rm M_{BH}\,{<}\, 10^7\,\msun$, the main distinctions appear at high-$z$ ($z\,{>}\,3$) where the \Wandering{} model predicts a smaller fraction of mass gained through gas accretion compared to the \Fiducial{} model. Although these differences are relatively small, there is a systematic trend. These high-$z$ differences increase towards larger masses and reach their maximum for objects with ${>}\,10^8\, \msun$. In this range, the \Fiducial{} model predicts a gas contribution of approximately 80\%, while the \Wandering{} one predicts about 70\%. At $z\,{<}\,2$ there are also some differences in MBHs with $\rm M_{BH}\,{>}\, 10^7\,\msun$. Specifically, the \Wandering{} model tends to predict larger gas contributions than the \Fiducial{} one. This is related to the trends seen in previous sections, where the off-center seed formation delays the assembly of MBHs and causes galaxies to quench later. As a result, galactic mergers or disc instabilities supply larger gas reservoirs around the MBHs compared to the model where seeds form in the galactic center.

In the right panel of Fig.~\ref{fig:Gas_Merger_accretion} we present the same analysis restricted to the active MBH population ($\rm L_{bol}\,{>}\,10^{44}\, erg/s$). The trends are similar to those presented above, but the active population features slightly smaller differences between the \Fiducial{} and \Wandering{} models. All these trends suggest that the off-center formation and subsequent wandering of the first seeds influence the amount of mass that MBHs gain through gas accretion, increasing the importance of mass acquired via MBH mergers. Despite this, we do not find any specific mass larger than $10^4\,\msun$, luminosity greater than $10^{44}\, \rm erg/s$, or redshift at which mergers surpass the mass gained from gas accretion. 

\subsection{The population of ungrown MBHs at high-$z$}

\begin{figure}    
\centering  
\includegraphics[width=1.\columnwidth]{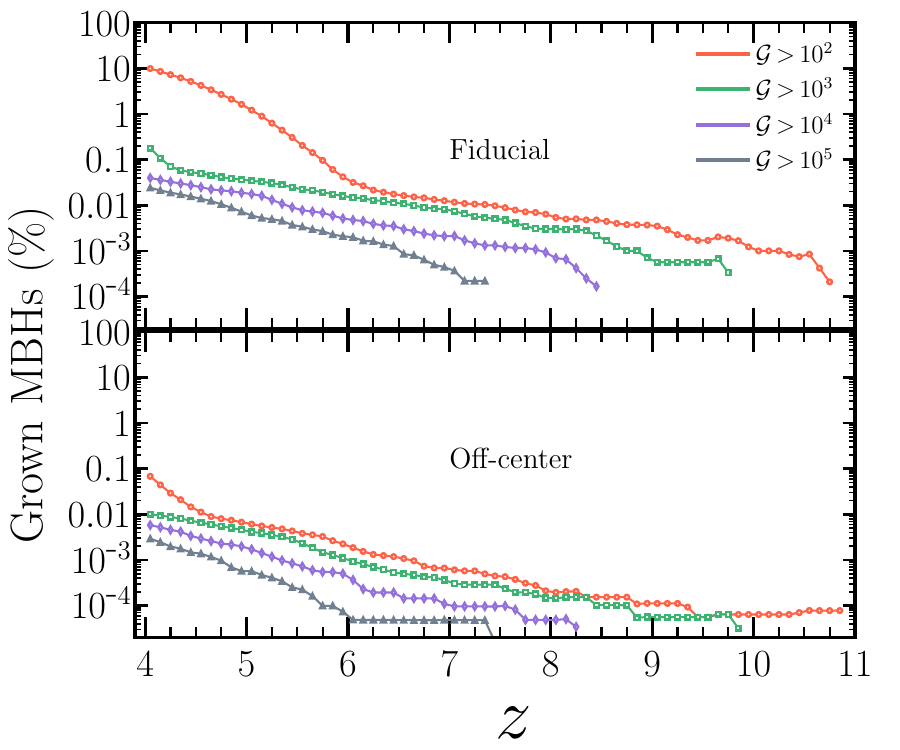}
\caption[]{Redshift evolution of the percentage of MBHs that grew at least a fraction $\mathcal{G}$ of its initial seed mass  (top and bottom: \Fiducial{} and \Wandering{} model, respectively). Different colors correspond to different $\mathcal{G}$ values. While upper panel corresponds to the \Fiducial{} model, the lower one represents the \Wandering{} case. 
}
\label{fig:Grown_MBHs}
\end{figure}

As discussed in the previous sections, off-center periods of MBH seeds lead to delays in the assembly of the MBH population. In this section, we examine the percentage of MBHs that can grow substantially at high-$z$. To do this, we will use the variable $\mathcal{G}$, which is defined as:
\begin{equation}
    \mathcal{G} \,{=}\, \rm \frac{M_{BH}(\mathit{z})}{M_{BH}(\mathit{z_f})}
\end{equation}
where $\rm M_{BH}(\mathit{z})$ is the mass of the MBH at the current redshift and $\rm M_{BH}(\mathit{z_f})$ the one corresponding to the formation time (i.e the seed mass). Fig.~\ref{fig:Grown_MBHs} presents the percentage of the MBH population that grew at least a value $\mathcal{G}$ above its initial seed mass. Notice that here, the full population consists of MBHs located at the centers of galaxies, as well as those that are off-center, wandering within the galaxy disc after a galaxy merger or off-center formation. Although both models show an increasing fraction of MBHs that have grown to a significant portion of their initial mass over time, there are notable differences between them. At $z\,{=}\,9$, the \Fiducial{} model predicts that only about $10^{-3}\%$ of the entire population has grown by a factor between $10^2\,{<}\,\mathcal{G}\,{<}\,10^3$. In contrast, at the same redshift, the \Wandering{} model finds an even lower value: less than $10^{-4}\%$. By redshift $z\,{=}\,7$, this fraction rises to approximately $0.01\%$ in the \Fiducial{} model, while in the \Wandering{} case remains as low as $2\,{\times}\,10^{-4}\%$. At $z\,{=}\,4$, the percentage varies significantly depending on the growth factor $\mathcal{G}$. For example, when $\mathcal{G}\,{>}\,10^2$ ($\mathcal{G}\,{>}\,10^3$), the \Fiducial{} model predicts that about $10\%$ ($0.1\%$) of the population has experienced a relevant growth. These numbers decrease considerably in the \Wandering{} model, where only about $0.1\%$ ($0.01\%$) of the population has grown by that factor.\\

The results presented here closely align with findings from previous sections, emphasizing that the off-center seed formation hampers the build-up of a significant population of intermediate-mass black holes in the early Universe ($z\,{>}\,4$). In fact, these small MBHs need to wait until later times to sink into the galactic center, where they can undergo substantial accretion episodes that significantly increase their masses. Similar conclusions were reached in the hydrodynamical simulation of \cite{Smith2018}, which showed that PopIII remnant seeds exhibit very inefficient growth at $9\,{<}\,z\,{<}\,15$. The authors reported that, on average, these MBH seeds increased their initial mass by a factor of $10^{-5}$, with the most active seeds growing by about 10\% \citep[see also][]{Liu2020}. Additionally, the study by \cite{Pfister2019} supports these findings. Specifically, they showed that high-$z$ MBHs with masses ${<}\,10^5 \msun$ are scattered throughout the galaxy due to irregularities in the gas and stellar potential. This causes them to oscillate around the galaxy center, keeping them away from dense gas regions and, thus, inhibiting significant accretion events.

\begin{figure}    
\centering  
\includegraphics[width=1.\columnwidth]{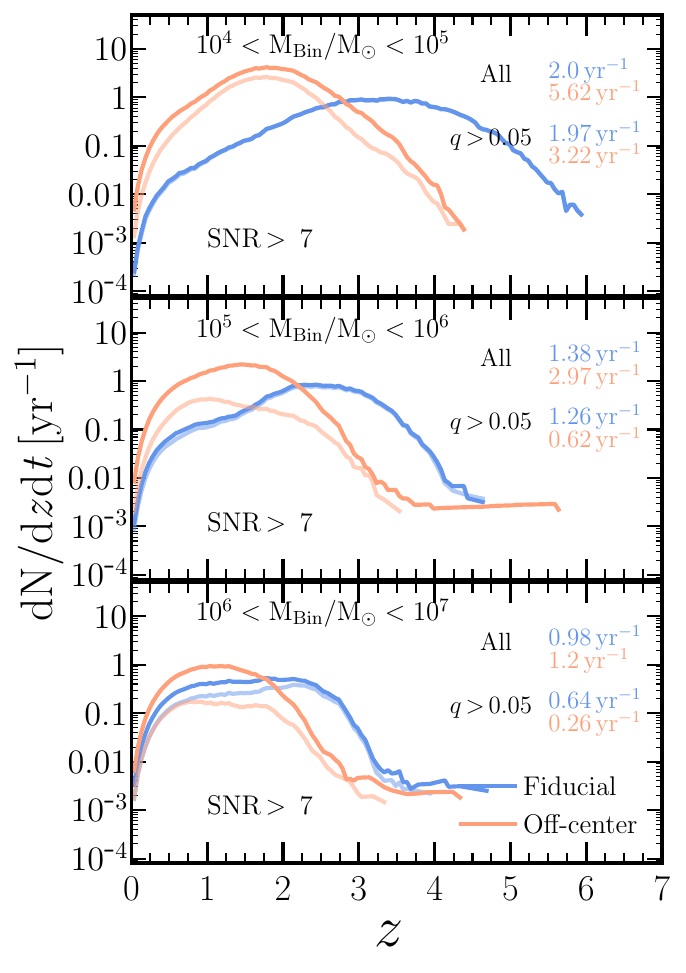}
\caption[]{MBHB merger rate detected by LISA with $\rm SNR\,{>}\,7$. Each panel corresponds to merged binaries with different total mass ($\rm M_{Bin}$). While blue lines correspond to the predictions of the \Fiducial{} model, orange ones represent the same but for the \Wandering{} one. Pale lines correspond to the merger rates related to merging binaries with mass ratio $q\,{>}\,0.05$ is assumed. The results have been produced with \LGalaxies{} applied in the \texttt{Millennium-II} simulation. The inserted numbers inside each panel correspond to the integrated merger rate with all and mass-cut ($q$) MBHs in a given mass bin.}
\label{fig:Merger_Rate_LISA}
\end{figure}
    
\begin{figure}    
\centering  
\includegraphics[width=1.\columnwidth]{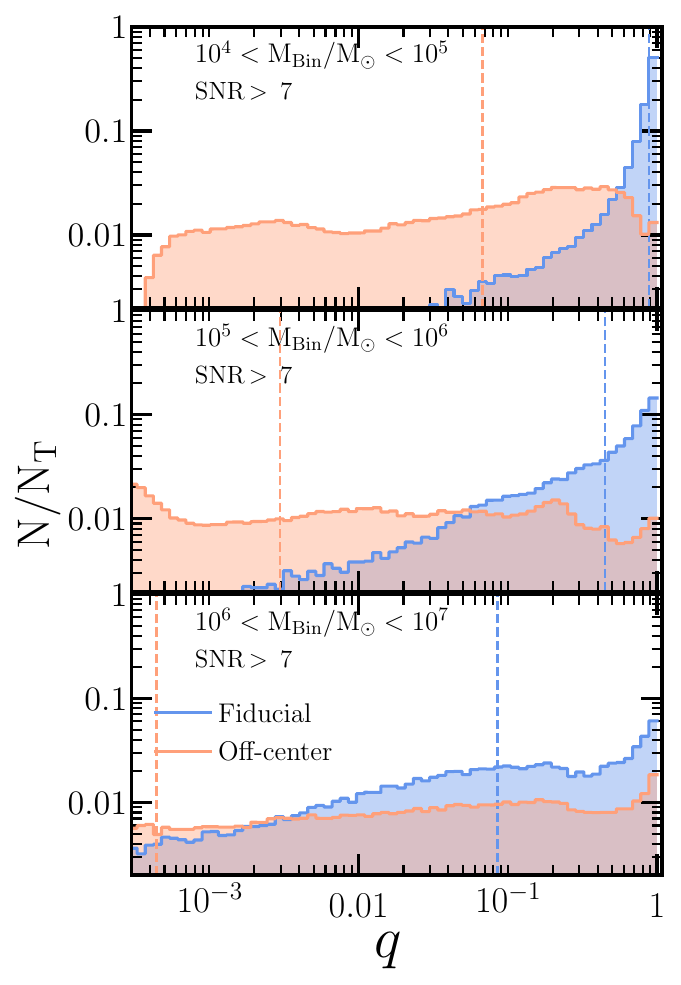}
\caption[]{Binary mass ratio, $q$, of merging MBHBs detected by LISA with $\rm SNR\,{>}\,7$. Each panel corresponds to merged binaries with different total mass ($\rm M_{Bin}$). While Blue histograms correspond to the predictions of the \Fiducial{} model, orange ones represent the same but for the \Wandering{} one. Horizontal dashed lines highlight the position of the median value of the distribution. The results have been produced with \LGalaxies{} applied in the \texttt{Millennium-II} simulation}
\label{fig:Mass_Ratio_LISA}
\end{figure}


\section{The population of MBHBs in the LISA band} \label{sec:Result_MBHBs}


The off-center formation of the first MBH seeds can also have a direct effect on the population of MBHBs. This section explores how binaries emitting GWs in the LISA band are affected. We stress that no relevant effects are seen in the PTA band (nHz) since \Fiducial{} and \Wandering{} model predict very similar MBH populations with masses ${>}\,10^8$ at $z\,{<}\,1$ (see Fig.~\ref{fig:BlackHoleMassFunction}).

The future LISA detector will detect the GWs emitted by the coalescence of MBHBs with $\rm 10^4\,{<}\,M_{BH}\,{<}\,10^{7}\,\msun$, a mass range affected by the off-center formation of the first MBH seeds (see Section~\ref{sec:Results_Nulcear_MBHs}). Thanks to the large sensitivity of the detector, it will be possible to assess whether the detections align more closely with the predictions of the MBH merger rates of the \Wandering{} or \Fiducial{} models. We caution however, that given the current uncertainties in the MBH formation and subsequent build-up it might be difficult (if at all possible) to single out the impact of off-center formation from the LISA data. 
Since \LGalaxies{} evolve a limited comoving volume of the universe, we can extract from the simulation the quantity $dn/dz$, representing the number of MBHB mergers per redshift and comoving volume. This can be used to compute the binary merger rate throughout the universe as
\begin{equation}
    \frac{d\mathrm{N}}{dzdt} \,{=}\, \left[\frac{4\pi \, c \, d_L^2}{(1+z)^2}\right] \left(\frac{dn}{dz}\right)\,,
\end{equation}
where $z$ is the redshift at which the merger occurs, $c$ the light speed and $d_L$ the source luminosity distance. Since we are interested in LISA detection rates, we count in $dn/dz$ only systems with signal-to-noise ratio ($\rm SNR$) larger than a threshold set to $7$ \citep[see][]{Colpi2024_RedBook}. 
The value of (sky-average) SNR assigned to each MBHBs will be computed as:
\begin{equation} \label{eq:SNR_equation}
    \rm SNR\,{=}\, \left(\frac{16}{5}\, \int_{\mathit{f}_0}^{\mathit{f_f}} \left[ \frac{\mathit{h}_c(\mathit{f}'_{obs})}{\mathit{h}_n(\mathit{f}'_{obs})}\right]^2 \frac{\mathit{df}'_{obs}}{\mathit{f}'_{obs}} \right)^{1/2},
\end{equation}
where $h_{\rm n}$ is the characteristic strain noise parameterized as in \cite{Babak2021} and  $h_{\rm c}$ the characteristic strain amplitude of the source. The latter is defined as $h_{\rm c}(f)\,{=}\,4f^2|\widetilde{h}(f)|^2$, being $\widetilde{h}(f)$ the Fourier transform of the strain signal, computed according to the phenomenological frequency-domain gravitational waveform model \textit{PhenomC} \citep{Santamaria2010}. The value of $f_0$ represents the starting frequency of inspiralling binaries, set randomly between $[0\,{-}\,T_{\rm obs}]$ being $T_{\rm obs}\,{=}\,4\,\rm yr$ the length of LISA observations. Instead, $f_f$ is the maximum frequency of the signal set to $0.15 c^3/G (1+\mathit{z}) \,\rm {M_{Bin}}$ where $\rm M_{Bin}$ is the total mass of the binary. The resolution frequency bin, $df$, used to integrate Eq.~\ref{eq:SNR_equation} is set to $df\,{=}\, 1/T_{\rm obs}$. Finally, the effective spin of the MBHBs is taken from the \LGalaxies{} predictions, and the eccentricity is set to $0$, for simplicity.\\

The predicted merger rates from the \Fiducial{} and \Wandering{} models are shown in Fig.~\ref{fig:Merger_Rate_LISA}. Generally, the overall merger rates for the former is $\rm {\sim}\,10 \, yr^{-1}$ while for the latter the rate is $\rm {\sim}\,5 \, yr^{-1}$. These differences are explained when the population is divided by mass.  For binaries with $\rm 10^4{<}\,M_{Bin}\,{<}\,10^5\, \msun$, the \Fiducial{} model predict that the first detected mergers start at $z\,{\sim}\,7$ (${\sim}0.01\,\rm yr^{-1}$), peak at $z\,{\sim}\,4$ (${\sim}\,0.7\, \rm yr^{-1}$), and quickly decline rapidly by $z\,{\sim}\,0$ (${\sim}10^{-3}\,\rm yr^{-1}$). In contrast, the \Wandering{} model presents a different picture. At $z\,{>}\,3$, mergers are significantly suppressed with rates that can be 5 to 10 times lower than those predicted by the \Fiducial{} case. At lower redshifts, the detection rate is boosted, peaking at $z\,{\sim}\,1.5$, where it is 10 times larger ($\rm {\sim}\, 4\,\rm yr^{-1}$) than the one of the \Fiducial{} model.  Interestingly, when the rates within this mass range are integrated over all redshifts, the \Wandering{} model predicts a value about 2.5 times higher than the \Fiducial{} case. For systems with $\rm 10^5{<}\,M_{Bin}\,{<}\,10^6\, \msun$ the scenario is very similar. While the \Fiducial{} model predicts that detected coalescences start at $z\,{\sim}\,5$ ($\rm {\sim}\, 0.1\,\rm yr^{-1}$) and peak at $z\,{\sim}\,3$ ($\rm {\sim}\, 0.8\,\rm yr^{-1}$), the \Wandering{} model shows an important suppression of high-$z$ mergers ($z\,{\geq}\,3$) in favour of an enhancement at $1\,{<}\,z\,{<}\,2$ ($\rm {\sim}\, 2\,\rm yr^{-1}$). As before, when the mergers are integrated over redshifts, the \Wandering{} model predicts a factor 2 larger events than the \Fiducial{} case. Finally, for MBHBs with $\rm 10^6{<}\,M_{Bin}\,{<}\,10^7\, \msun$ the picture is similar, although the integrated merger rates of the two models are comparable in this case.\\

As described above, the differences between the \Fiducial{} and \Wandering{} models can be summarised in two main features: a suppression of events at high-$z$ and an enhancement at low-$z$. The first trend results directly from the off-centre seed formation, which causes only a few high-$z$ galaxies to host nuclear MBHs. Consequently, when MBHs, following a galaxy merger and a dynamical friction phase, reach the centre of their galaxies, very few of these encounter another nuclear MBH to form a binary system. The second trend is the result of a combination of the delayed MBH growth and the multi-PopIII seeding introduced in Section~\ref{sec:WanderingModel}. On the one hand, the delayed assembly of MBHs results in longer sinking timescales for the population, meaning they spend extended periods in the dynamical friction phase after a galaxy merger. As a result, the formation and the eventual merger of MBHBs take longer times in the \Wandering{} model than in the \Fiducial{} one. On the other hand, the multi-PopIII seeding in the \Wandering{} model leads to galaxies hosting more light seeds that wander within the galaxy compared to the \Fiducial{} model. As a result, some of these seeds can reach the galaxy centre at low-$z$ and boost the merger rate. Interestingly, these two last points imply a lighter MBHB population reaching the galactic nucleus. This is evident in Fig.~\ref{fig:Mass_Ratio_LISA}, which shows the mass ratio, $q\,{=}\,\rm M_{BH,2}/M_{BH,1}$, of LISA detected MBHBs. As illustrated, for systems with total mass $\rm 10^4{<}\,M_{Bin}\,{<}\,10^5\, \msun$, the \Fiducial{} model predicts mergers with nearly equal masses, with a median $q\,{\sim}\,1$. In contrast, the $q$ values in the \Wandering{} scenario extend down to $10^{-3}$ (i.e the range corresponding to intermediate-mass-ratio inspirals, IMRIs), with a median value approximately ten times smaller ($q\,{\sim}\,0.1$) than in the \Fiducial{} case. Similar trends are seen in the other two mass bins explored. For the case of $\rm 10^5{<}\,M_{Bin}\,{<}\,10^6\, \msun$ ($\rm 10^6{<}\,M_{Bin}\,{<}\,10^7\, \msun$) the \Fiducial{} model finds mergers with typical $q\,{\sim}\,0.5$ ($q\,{\sim}\,0.09$) while the  \Wandering{} case is $q\,{\sim}\,3\,{\times}\,10^{-3}$ ($q\,{\sim}\,7\,{\times}\,10^{-4}$). \\


As shown, in Fig.~\ref{fig:Mass_Ratio_LISA}, a considerable number of the mergers in the \Wandering{} model involve systems with mass ratios small enough to be consistent with IMRIs. Currently, accurate waveform models for $q\,{\lesssim}\,0.05$ are not yet fully developed \citep[see e.g][]{Babak2017,Hinder2018}, which limits our ability to precisely characterise the merging signals of these systems. Therefore, it is more appropriate to focus on merger rates where the mass ratio exceeds this threshold, ensuring the reliability of waveform modelling and parameter estimation\footnote{It is also important to note that if the SNR is sufficiently high, the detection of IMRIs may still be feasible even with less accurate waveform templates \citep{Babak2017}. Nonetheless, the primary challenge lies in the potential biases introduced in parameter estimation if waveform models are not sufficiently accurate for these systems \citep{Hinder2018}. This underscores the critical need for developing precise IMRI waveform models, which will enhance both detection prospects and the robustness of astrophysical inferences from future GW observations.}. 
The results with $q\,{>}\,0.05$ are shown with the pale lines in Fig.~\ref{fig:Merger_Rate_LISA}. As expected, the merger rates of the \Fiducial{} model remain unaffected by this $q$-value cut. In contrast, the \Wandering{} model is significantly impacted, especially in systems with higher total mass. Overall, the merger rates decrease by a factor of 3 to 9, resulting in the \Fiducial{} model predicting a higher number of mergers than the \Wandering{} model.

To conclude, the LISA experiment might have the potential to reveal and constrain the physical process of off-centre seed formation. If this process plays a significant role in the evolution of the first MBHs, LISA is expected to detect fewer nearly equal-mass mergers than initially anticipated. Furthermore, these would mainly occur at low-$z$. However, it is worth noticing that these features might not be unique to off-center seeding and other physical processes could potentially lead to similar observational signatures. 

\section{Conclusions} \label{sec:Conclusions}
By using the state-of-the-art semi-analytical model \LGalaxies{} applied on the \texttt{Millennium} suite of simulations, we have studied how off-center MBH seed formation affects the global population of MBHs and MBHBs. To this end, we have extended the regular galaxy formation model of \LGalaxies{} (\Fiducial{} model) assuming that light seeds (PopIII remnants) do not form in the galactic center but distributed in the minihalo where they are born. This implies that light seeds do not meet ideal conditions to grow by mergers and gas accretion at birth, which causes a delay in their mass evolution (\Wandering{} model). The main findings of this work can be summarised as follows:
\begin{itemize}
    \item The off-center seed formation has an important effect on the assembly of the MBH population. At high-$z$ ($z\,{>}\,4$) the abundances of MBHs with ${>}\,10^5\,\msun$ are suppressed by a factor 2-10 compared to the \Fiducial{} model. This implies that at these redshifts, less than 0.01\% of the entire MBH population manages to grow. At lower-$z$ ($z\,{<}\,2$), the differences in the ${>}\,10^5\,\msun$ MBH population have almost vanished since the predictions of the \Wandering{} and \Fiducial{} models converge to similar results.\\
    
    \item Quasar AGN luminosity functions are affected by off-centre seed formation. At $z\,{>}\,6$, the \Wandering{} model predicts five times fewer systems with $\rm L_{bol}\,{>}\,10^{44} \, erg/s$ compared to the \Fiducial{} model. At lower redshifts, the two models generally converge, although the \Wandering{} model predicts slightly more sources. This is because the delayed assembly of MBHs results in less effective galaxy quenching. As a result, galaxies have more gas available to feed MBHs, allowing them to stay active for a longer period.\\ 

    \item Important differences are seen in $\rm M_{Stellar}\,{-}\,M_{BH}$ correlation. At $z\,{>}\,5$ the amplitude predicted by the \Wandering{} model is lower than the \Fiducial{} one, with MBHs up to $2\,\rm dex$ smaller at a fixed stellar mass. At $z\,{\sim}\,2$ this difference vanishes for galaxies with $\rm M_{Stellar}\,{>}\,10^{11}\, \msun$ but it persists in systems with smaller mass. In the local Universe, the differences between the two models vanish, and no signature of the off-center formation is left in the scaling relation. On top of this, the overmassive population of MBHs recently discovered by JWST is present in both \Fiducial{} and \Wandering{} model and shows very similar $\rm M_{Stellar}\,{-}\,M_{BH}$ scaling relation. Thus, the correlation of this population with their host cannot constrain any off-center formation of MBHs.\\ 

    \item The rate of merging MBHs within the LISA band ($\rm 10^4\,{<}\,M_{BH}\,{<}\,10^7\, \msun$) is significantly influenced by the off-center seed formation. This scenario leads to a suppression of events at high redshift and an enhancement at low redshift compared to the \Fiducial{} model. The \Wandering{} model exhibits a higher overall merger rate ($\rm {\sim}\,10\, yr^{-1}$) compared to the \Fiducial{} model ($\rm {\sim}\,5\, yr^{-1}$). This difference is primarily driven by a larger number of IMRIs in the \Wandering{} model. However, when applying a mass ratio threshold of ${>}\,0.05$, the \Fiducial{} model yields a greater number of detectable events. These trends suggest that LISA could potentially constrain off-centre seed formation and the early dynamical evolution of MBHs. Nonetheless, such features may not be unique to off-center seeding, as other physical processes could potentially lead to similar observational signatures. 
\end{itemize}

The results presented in this work represent the first attempt to statistically assess the effects of off-center seed formation and the subsequent wandering phase on the population of MBHs and MBHBs. Despite that, further investigation is needed. A synergy between semi-analytical models and the outcomes of hydrodynamical simulations that do not reposition the MBH at the center of the galaxy \citep{Dubois2014,Tremmel2017,Dubois2021,Trebitsch2021A} would pave the path to a better understanding of the dynamical evolution of the first seeds within galaxies. 

\begin{acknowledgements}
We thank the B-Massive group at Milano-Bicocca University for useful discussions and comments. D.I.V. thanks John Wise for his useful discussion about the birthplace of the first light seeds. D.I.V and A.S. acknowledge the financial support provided under the European Union’s H2020 ERC Consolidator Grant ``Binary Massive Black Hole Astrophysics'' (B Massive, Grant Agreement: 818691) and the European Union Advanced Grant ``PINGU'' (Grant Agreement: 101142079). M.V. acknowledges funding from the French National Research Agency (grant ANR-21-CE31-0026, project MBH\_waves) and from the Centre National d’Etudes Spatiales. M.C. acknowledges funding from MIUR under the grant PRIN 2017-MB8AEZ, from the INFN TEONGRAV initiative, and from the MUR Grant "Progetto Dipartimenti di Eccellenza 2023-2027” (BiCoQ). D.S. acknowledges support by the Fondazione ICSC, Spoke 3 Astrophysics and Cosmos Observations. National Recovery and Resilience Plan (Piano Nazionale di Ripresa e Resilienza, PNRR) Project ID CN\_00000013 "Italian Research Center on High-Performance Computing, Big Data and Quantum Computing" funded by MUR Missione 4 Componente 2 Investimento 1.4: Potenziamento strutture di ricerca e creazione di "campioni nazionali di R\&S (M4C2-19 )" - Next Generation EU (NGEU). S.B. acknowledges support from the Spanish Ministerio de Ciencia e Innovación through project PID2021-124243NB-C21. M.V. acknowledges funding from the French National Research Agency (grant ANR-21-CE31-0026, project MBH\_waves) and from the Centre National d’Etudes Spatiales.
\end{acknowledgements}

%
%

\bibliographystyle{aa.bst} 
\bibliography{references.bib} 

\appendix

\section{The formation places of MBH seeds and their wandering times} \label{Appendix:R0_and_Twand}

\begin{figure}[h]   
\centering  
\includegraphics[width=1\columnwidth]{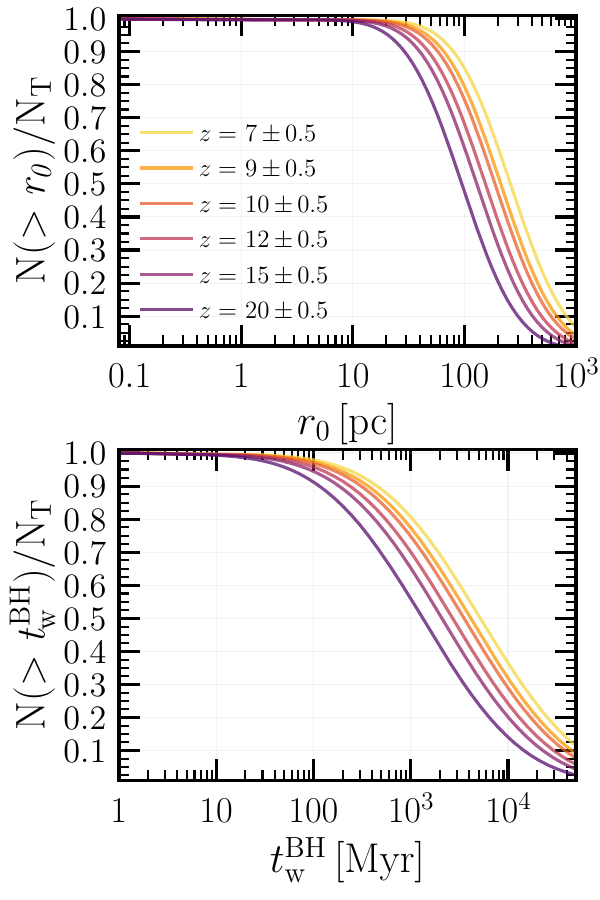}
\caption[]{Cumulative distribution function of the initial position ($r_0$, upper panel) and wandering time ($t_{\rm w}^{\rm BH}$, lower panel) of newly formed seeds from the run of \LGalaxies{} on the  \texttt{Millennium-II} merger trees. Each color represents a different redshift bin.}
\label{fig:Distance_and_Wandering_Time}
\end{figure}

The upper panel of Fig.~\ref{fig:Distance_and_Wandering_Time} shows the cumulative distribution of $r_0$ for PopIII remnants at different redshifts. At $z\,{>}\,15\,$(7) around 50\% of the seed population is placed at ${>}\,100\,(300) \,\rm pc$ from the galactic centre. This redshift evolution is just the natural result of halos being more extended towards lower redshift. As a result, the $r_0$ value of an MBH seed formed at low-$z$ will be extracted from a wider range of distances than a seed born at higher redshifts.\\

To show the typical time spent by MBH seeds in the wandering phase, the lower panel of Fig.~\ref{fig:Distance_and_Wandering_Time} shows the cumulative distribution function of $t_{\rm w}^{\rm BH}$. As shown, 50\% (20\%) of the newly formed MBH seeds at $z\,{>}\,15$ is expected to undergo a wandering phase that lasts ${>}\,2 \, \rm Gyr$ (${<}\,100\, \rm Myr$). At lower redshifts, the duration of this phase increases. For example, 50\% (20\%) of the seeds formed at $z\,{\sim}\,7$ will wander for ${>}\,4 \, \rm Gyr$ (${<}\,400\, \rm Myr$).

\end{document}